\documentclass{article}
\usepackage{spconf,amsmath, amssymb, graphicx}
\usepackage{eufrak}
\usepackage{amsthm}
\usepackage{booktabs}
\usepackage{csquotes}
\usepackage{subfig}
\usepackage{float}
\usepackage{bm}
\usepackage{siunitx}
\usepackage{multirow}
\usepackage{hyperref}

\def\m1{\mathbf{m}_1}
\def\m2{\mathbf{m}_2}

\newcommand{\norm}[1]{\left\lVert#1\right\rVert}

\title{MIRAGE: 2D SOURCE LOCALIZATION USING\\ MICROPHONE PAIR AUGMENTATION WITH ECHOES}

\name{Diego Di Carlo$^\dagger$, Antoine Deleforge$^\ddagger$, and Nancy Bertin$^\dagger$\thanks{The research presented in this paper is reproducible. Code and data are available at \url{https://github.com/Chutlhu/MIRAGE}}}
\address{$^\dagger$ Univ Rennes, Inria, CNRS, IRISA, France\\
         $^\ddagger$ Universit\'e de Lorraine, CNRS, Inria, LORIA, F-54000 Nancy, France}
\begin{document}
%\ninept

\maketitle

\begin{abstract}
It is commonly observed that acoustic echoes hurt performance of sound source localization (SSL) methods. We introduce the concept of microphone array augmentation with echoes (MIRAGE) and show how estimation of early-echo characteristics can in fact benefit SSL. We propose a learning-based scheme for echo estimation combined with a physics-based scheme for echo aggregation. In a simple scenario involving 2 microphones close to a reflective surface and one source, we show using simulated data that the proposed approach performs similarly to a correlation-based method in azimuth estimation while retrieving elevation as well from 2 microphones only, an impossible task in anechoic settings.
\end{abstract}
\section{Introduction}
\label{sec:intro}
% \textbf{2D SSL enhanced by Echoes}
Sound source localization (SSL) consists in determining the position of a sound source from microphone signals in 3D space. In polar coordinates, most existing methods focus on estimating the directional of arrival, namely, azimuth and elevation angles. Though this task is performed routinely by humans, it still challenges today's computational methods, in particular in the presence of reverberation or interfering sources (see \cite{rascon2017localization} and \cite{Argentieri2015} for a review). Computational approaches consist in two components. First, extracting features from audio data that are as independent as possible from the source's content while preserving spatial information. Second, mapping these features to the source position. Two lines of research have been investigated to obtain such mappings: physics-based and learning-based approaches.

\begin{figure}[!t]
    \centering
    \includegraphics[trim={50 70 50 150},clip,width=\linewidth]{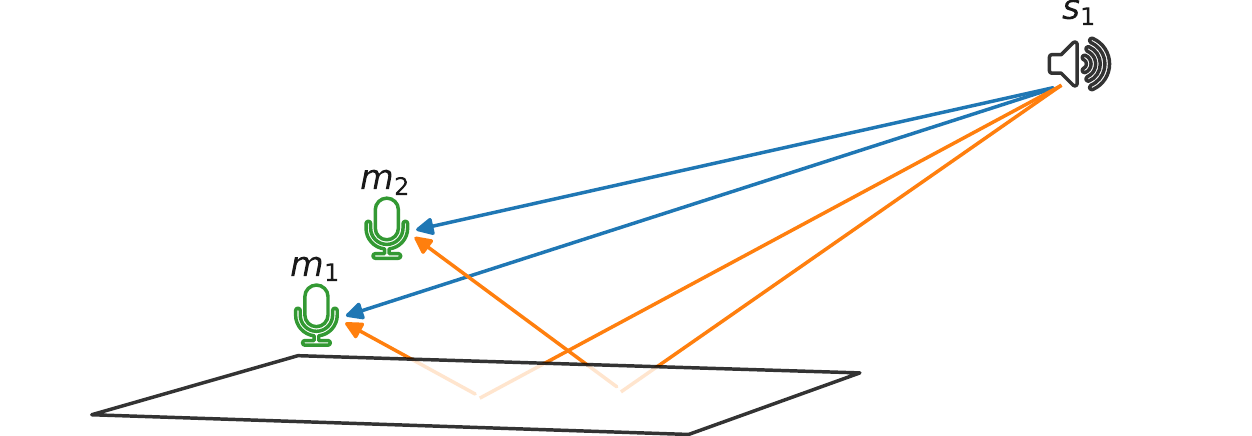}
    \caption{\label{fig:scene}Typical setup with one source source recorded by two microphones. The illustration shows direct sound path (blue lines) and resulting first-order echoes (orange lines).    \vspace{-5mm}}
\end{figure}

Physics-based approaches rely on a simplified sound propagation model \cite{rascon2017localization,Knapp1976,DiBiase2001,Lebarbenchon2018}. The free-field model is by far the most widely used one and assumes a single direct sound path from the source to each microphone. When the source is placed far enough, this yields a closed-form mapping from the sound's time-difference-of-arrival (TDOA) in a microphone pair and the source's azimuth angle in this pair. If multiple microphone pairs are available and form a non-linear array, their TDOAs can be aggregated to obtain 2D directions of arrival \cite{DiBiase2001}. These methods strongly suffer in environments where the free-field assumption is violated, \textit{e.g.}, in the presence of strong acoustic echoes and reverberation \cite{Scheuing2006}.

Learning-based approaches use an annotated training dataset to implicitly learn a mapping from audio features to source positions \cite{deleforge2015acoustic, Vesperini2016, Adavanne2017,  Perotin2018, gaultier2017vast}. Such data can be obtained from real recordings \cite{deleforge2015acoustic} or using physics-based simulators \cite{Vesperini2016, Adavanne2017,  Perotin2018, gaultier2017vast}. These methods were showed to overcome some limitations of the free-field model, but are usually trained for specific microphone arrays and fail whenever test conditions strongly mismatch training conditions.

Most sound source localization methods, including the above listed, regard reverberation and in particular acoustic echoes as a nuisance. In contrast, some recent work that we refer to as \textit{echo-aware} methods have showed that the knowledge of early acoustic echoes could be used to reconstruct the geometry of an audio scene \cite{Nakashima2010,dokmanic2013acoustic,An2018} or to improve performance of signal enhancement methods \cite{flanagan1993spatially, dokmanic2015raking,Scheibler2017}. In \cite{Nakashima2010}, some ad-hoc reflectors are used as artificial \textit{pinnae} to estimate elevation based on a simple reflection model. In \cite{An2018}, cameras, depth sensors and laser sensors are used to identify reflectors and build a corresponding acoustic model that helps SSL.

In this work, we combine ideas from physics-based, learning-based and echo-aware approaches to introduce the framework of \underline{mi}crophone a\underline{r}ray \underline{a}u\underline{g}mentation with \underline{e}choes (MIRAGE) for SSL. We consider a simple yet common scenario to illustrate this idea: two microphones, one source and a nearby reflective surface, as illustrated in Fig. \ref{fig:scene}. This may occur, for instance, when the sensors are placed on a table such as in voice-based assistant devices or next to a wall. The reflective surface is assumed to be the most reflective and closest one to the microphones in the environment, hence generating the strongest and earliest echo in each microphone. Under this \textit{close-surface} model, we ask the following questions:
\begin{enumerate}
\setlength\itemsep{-1mm}
\item Can early echoes be estimated from two-microphone recordings of an unknown source?
\item Can they be used to estimate both the azimuth and elevation angles of the source, an \textit{impossible} task in free field conditions?
\end{enumerate}
We propose to use a deep neural network (DNN) trained on a simulated close-surface dataset to estimate early echoes properties from audio features. The MIRAGE framework then exploits these estimated properties by expressing them as TDOAs in the \textit{virtual 4-microphone array} formed by the true microphone pair and its image with respect to the reflective surface. We show that the proposed framework approximately estimates echo properties, perform similarly to a correlation-based method in azimuth estimation for the considered scenario and estimates \textit{impossible} elevation angles with good accuracy in noiseless settings using two microphones only.

\section{Background in microphone array SSL}\label{sec:background}
In this section, we briefly review some necessary background in microphone array SSL. Let us assume a microphone array of $I$ sensors is placed inside a room and records the sound emitted by one static point sound source. In all generality, the relationship between the signal $m_i(t)$ recorded by the sensor placed at fixed position $\mathbf{m}_i$ and the signal $s(t)$ emitted by the source at fixed position $\mathbf{s}$ is defined by:
\begin{equation}\label{eq:anymic_time}
m_i(t) = (h_i * s)(t)  \; + \; n_i(t),
\end{equation}
where the convolution with room impulse response (RIR) $h_i(t)$ embodies the fact that sensor $i$ receives a so-called spatial image of the source and $n_i$ denotes possible measurement noise. The RIR depends on the spatial parameters of the scene: microphone positions, source position w.r.t the room, as well as the room acoustic properties (size, absorption and diffuseness of the wall materials.)

\begin{figure}
    \centering
    \includegraphics[trim={0 5 0 0},clip,width=\linewidth, height=0.6\linewidth]{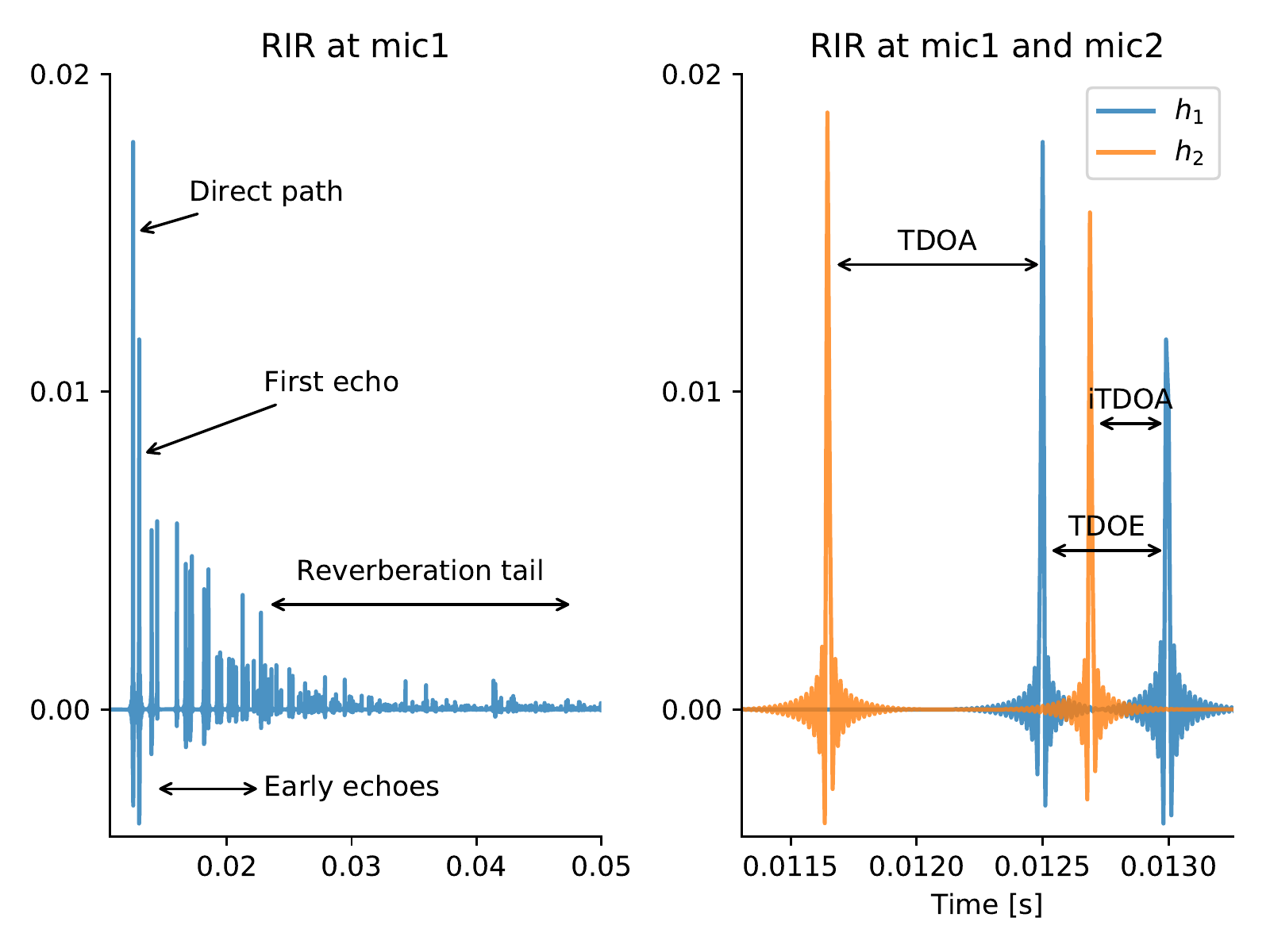}
    \caption{\label{fig:rirs}Left, a typical simulated RIR with annotated components. Right, superposition of two RIRs and visualization of time difference of arrival between direct paths (TDOA), first echoes (iTDOA) and direct path and first echo (TDOE). \vspace{-4mm}}
\end{figure}

RIRs can be typically modelled as the sum of the direct path and multiple reflections of the sound. This can boil down to modelling $h_i$ as a Dirac impulse at time $\tau_i$ accounting for the time delay from the source to microphone $i$, plus an error term. In the frequency domain, this leads to:
\begin{equation}\label{eq:rir}
H_i(f) = \alpha_i(f) \; e^{- 2 \pi f \tau_i} \; + \; \varepsilon_i(f),
\end{equation}
where the error term $\varepsilon_i(t)$ collects echoes, the reverberation tail, diffusion, and noise. The term $\alpha_i(f)$ captures the air attenuation phenomenon. A time-domain example of RIR is shown in Fig.~\ref{fig:rirs} (left).

%A strong relationship exists between the values of delays $\tau_i$ and the source direction of arrival (DOA), i.e. the azimuth and elevation angles $(\theta, \phi)$ of the source in the microphone array frame, centered on its barycenter. Under favorable conditions, estimating the source direction is possible when several microphones are available

\subsection{2-channel 1D-SSL}
\label{subsec:1D-SSL}
Let us first consider the stereo case ($I=2$). Under the far-field assumption, traditional SSL methods use the time difference of arrival (TDOA), $\tau \triangleq \tau_2 - \tau_1$, as a proxy for the estimation of the angle of arrival (AOA), since:
\begin{equation}\label{eq:aoa}
%\text{AOA} = \text{arccos} \left( \frac{c \: \tau}{d} \right),
\text{AOA} = \text{arccos} \left(c \: \tau \: / \:d \right),
\end{equation}
where $c$ is the speed of sound and $d$ the inter-microphone distance. SSL then reduces to estimating the TDOA, which can be done by cross-correlation-based methods such as %between the two microphones signals, such as 
the widely used and well performing generalized cross-correlation with phase transform (GCC-PHAT) method \cite{Knapp1976,Blandin2012}. Given short-time Fourier transforms $M_1$ and $M_2$ of the two microphones signals, the GCC-PHAT \textit{angular spectrum} is defined as: 
\begin{equation}\label{eq:gccphatcontrast}
\Psi_\text{GCC}(\tau) = \sum_{f,n}\frac{M_1(f,n) M_2^*(f,n)}{\mid M_1(f,n) M_2^*(f,n) \mid} e^{-2\pi f \tau}.
\end{equation}
Then, the TDOA estimate is given by $\hat{\tau} = \arg \underset{\tau}{\max} \; \Psi_\text{GCC}(\tau)$. Note that $\Psi_\text{GCC}$ can also be expressed directly as a function of the AOA using \eqref{eq:aoa}, hence the term \textit{angular spectrum}. %This method was showed to be state-of-the-art in a large benchmark \cite{Blandin2012}.

%where $gcc_\text{PHAT}$ is the inverse Fourier Transform of $GCC_\text{PHAT}$ . This method generalizes the cross-correlation method by normalizing the cross-spectral density in the frequency domain. 

%TDOA-based methods such as GCC-PHAT essentially ignore noise and echoes and assume that the direct paths are sufficiently prominent in the RIRs to enable correct estimation despite reverberation. It is also important to notice that the AOA computed form $\hat{\tau}$ is a \textit{local} azimuth in the coordinate system defined by the microphones pairs (1D-SSL). Back in 3D, it only locates the source up to a so-called ``cone of confusion'', making impossible to provide a full estimation of the global DOA ($\theta,\phi$). In particular, if the microphone pair is parallel to the global horizontal plane, AOA corresponds to azimuth $\theta$ or to its opposite (front-back ambiguity), and elevation $\phi$ cannot be estimated.

\subsection{Multichannel 2D-SSL}
\label{subsec:2D-SSL}
When more microphones are available and the array is not linear, 2D-SSL can be envisioned. A possible approach is to use 1D-SSL on all pairs and combine their results, a principle which was successfully applied in the steered response power with phase transform (SRP-PHAT) method \cite{DiBiase2001}. SRP-PHAT exploits the geometry of the microphone array and the estimated TDOAs from microphone pairs to return the DOA. In a nutshell, this algorithm aims to estimate a global angular spectrum $\Psi_{\text{SRP}}(\theta,\phi)$ which will exhibit a local maximum in the direction of the active source. First, a global grid of possible DOAs is defined according to a desired resolution and computational load. Second, for each pair of microphones, a local set of AOAs is defined and a TDOA-based algorithm (e.g. GCC-PHAT) is used to compute the associated local angular spectrum. Finally all the local contributions (a collection of local $\Psi_\text{GCC}(\tau)$) are geometrical aggregated and interpolated back to the global DOA grid to form $\Psi_{\text{SRP}}(\theta,\phi)$, and the DOA maximizing $\Psi$ is used as estimate.

%Again, a limitation of this method is to consider only the contribution of the direct paths, through TDOA; it does not aim at predicting or exploiting time locations of first echoes, which convey information, and can even be disturbed by their presence. More obviously, this method only makes sense for elevation estimation if strictly more than two microphones (one pair) are available; although it collapses to GCC-PHAT itself in the stereo case, the angular spectrum would then be constant wrt elevation.

%In the rest of this paper, we examine how an estimation of additional angular spectra for image microphone pairs can allow for iTDOA, TDOE1, TDOE2 (see Fig.~\ref{fig:scene}) and elevation estimations, as well as an improved azimuth estimation, even in the stereo case, by taking explicitly into account the first echoes in the RIRs.

\begin{figure}
    \centering
    \includegraphics[trim={90 75 40 50},clip,width=0.7\linewidth]{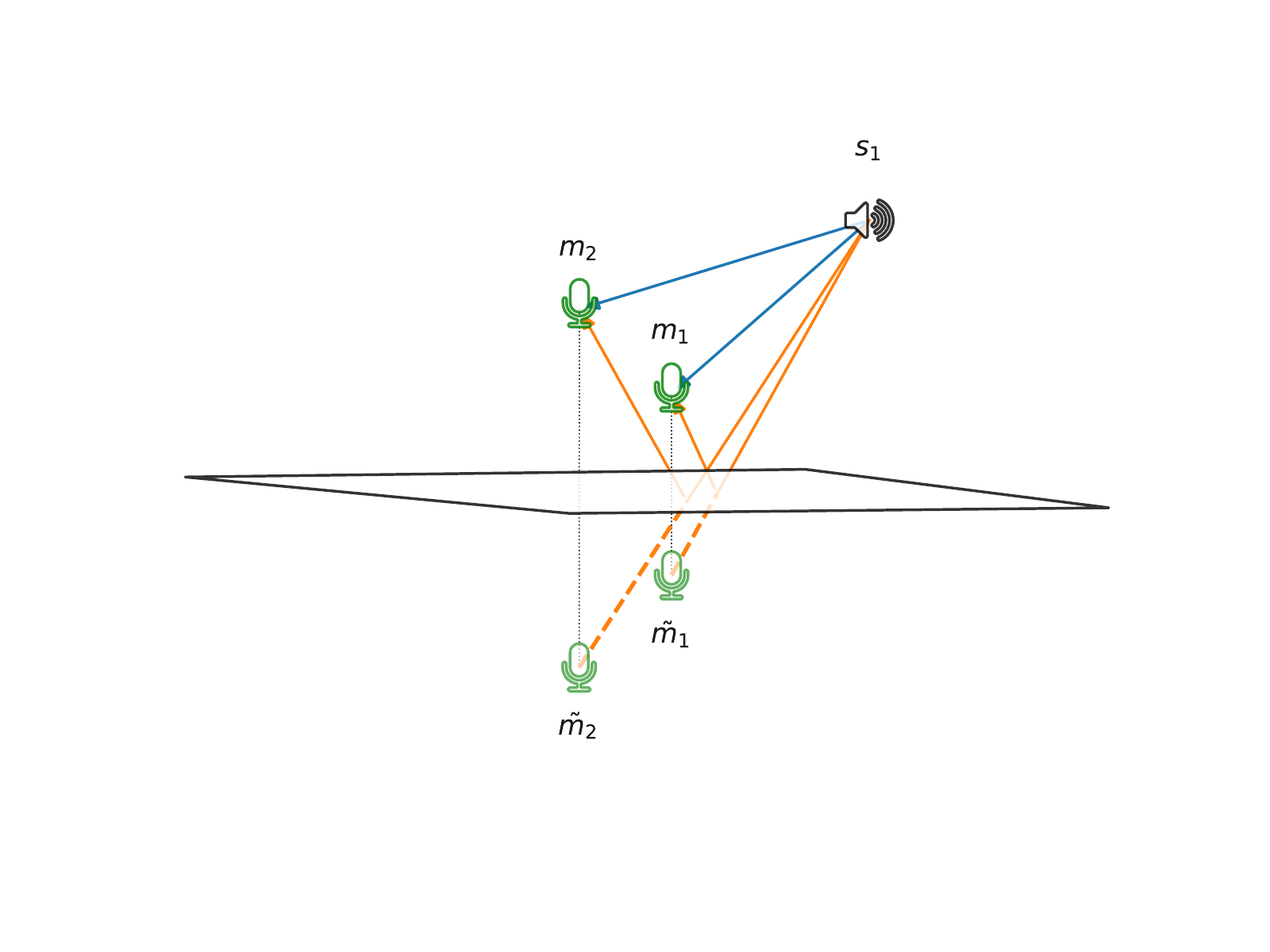}
    \caption{\label{fig:mirage}Illustration of the images $\tilde{m}_1$ and $\tilde{m}_2$ of microphones $m_1$ and $m_2$ in the presence of a reflective surface and a source. Blue lines correspond to direct paths, orange lines correspond to echo paths.\vspace{-4mm}}
\end{figure}

\section{MIRAGE: Microphone Array Augmentation with Echoes}\label{sec:mirage}
We now introduce the proposed concept of \underline{mi}crophone a\underline{r}ray \underline{a}u\underline{g}mentation with \underline{e}choes (MIRAGE).
Let us first expand formula~\eqref{eq:rir} to account for more echoes:
\begin{equation}
\label{eq:echo_h}
H_i(f) = \sum_{k=0}^{K}\alpha_i^k(f) \; e^{- 2 \pi f \tau_i^k} \; + \; \varepsilon_i(f)
\end{equation}
where the sum now comprises the direct path ($k=0$) and the $K$ earliest reflections ($K = 1$ in this paper) and $\varepsilon_i$ collects the remaining RIR components.
%Without loss of generality we can fix the delay of one direct path to zero, say $\tau_1^0 = 0$ for microphone $i=1$.
Here, $\alpha_i^k(f)$ accounts for both air attenuation and wall absorption phenomena. In the remainder of this paper, we make the approximation of frequency-independent $\alpha_i^k$. Eq. \ref{eq:echo_h} then corresponds to the well known image-source (IS) model, where reflections are treated as mirror images of the true source with respect to reflective surfaces, emitting the same signal. We will employ here a less common but equivalent interpretation of IS, namely, the image-microphone (IM) model. As illustrated in Fig.~\ref{fig:mirage}, virtual microphones are mirror images of the true microphones with respect to reflective surfaces. In this view, the echoic signal received at a true microphone is the sum of the anechoic signals received at this microphone and its images. If we consider the virtual array consisting of both true and image microphones, multiple microphone pairs are now available. For each of them, it is then possible to define a corresponding time difference of arrival. Among them, we will refer to the one between the two real microphones as TDOA, the one between the two image microphones as image TDOA (iTDOA) and the one between the first microphone and its image as time difference of echoes (TDOE). We have:
\begin{align}
%\text{TDOA} &= \dfrac{\norm{\mathbf{m}_2 - \mathbf{s}} 
%                - \norm{\mathbf{m}_1 - \mathbf{s}}}{c} 
%                = \tau_2^0 - \tau_1^0,\\
%\text{iTDOA}&= \dfrac{\norm{\tilde{\mathbf{m}}_2 - \mathbf{s}}
%                - \norm{\tilde{\mathbf{m}}_1 - \mathbf{s})}}{c}
%                = \tau_2^1 - \tau_1^1,\\
%\text{TDOE} &= \dfrac{\norm{\tilde{\mathbf{m}}_1 - \mathbf{s}}
%                - \norm{\mathbf{m}_1 - \mathbf{s}}}{c} 
%                = \tau_1^1 - \tau_1^0,
\text{TDOA} &= (\norm{\mathbf{m}_2 - \mathbf{s}} 
                - \norm{\mathbf{m}_1 - \mathbf{s}})/c 
                = \tau_2^0 - \tau_1^0,\\
\text{iTDOA}&= (\norm{\tilde{\mathbf{m}}_2 - \mathbf{s}}
                - \norm{\tilde{\mathbf{m}}_1 - \mathbf{s}})/c
                = \tau_2^1 - \tau_1^1,\\
\text{TDOE} &= (\norm{\tilde{\mathbf{m}}_1 - \mathbf{s}}
                - \norm{\mathbf{m}_1 - \mathbf{s}})/c 
                = \tau_1^1 - \tau_1^0,
\end{align}
where $\tilde{\mathbf{m}}_i$ denotes the image of position $\mathbf{m}_i$. These three quantities are directly connected to RIRs, as illustrated in Fig.~\ref{fig:rirs}(right). 
Let $V = \{ \text{TDOA}, \text{iTDOA}, \text{TDOE}\}\in\mathbb{R}^3$. Following the 2D-SSL scheme described in Sec. \ref{subsec:2D-SSL} and given the virtual microphone-array geometry (which depends on the relative position of microphones to the surface), $V$ could in principle be used to estimate the 2D directional of arrival of the source. In the next section, we present a learning-based method to estimate $V$ using audio features obtained from only two microphones. 

\section{Learning-based echo estimation}
% Our aim is to recover the time location of the direct path and the first echoes of the RIRs. Thus, it follow naturally to model this problem as a multi-target regression. 
Our approach is to train a deep neural network (DNN) on a dataset simulating the considered close-surface scenario. We model the problem as multi-target regression, with \textit{interaural level difference} (ILD) and \textit{interaural phase difference} (IPD) as input features, and $V \in \mathbb{R}^3$ as output parameters.
%Contrary to classification approaches previously used for DNN-based SSL [REF], this approach is off-grid and do not introduce any additional error due to the sampling and binning in the target space.
ILD and IPD features are defined in the frequency domain as follows:
\begin{equation}
\label{eq:features}
\begin{cases}
ILD(f)  =& \tfrac{1}{T} \sum_{t=1}^T \log{\mid \frac{M_2(f,t)}{M_1(f,t)} \mid } \\
IPD(f)  =& \tfrac{1}{T} \sum_{t=1}^T \frac{M_2(f,t)/ \mid M_2(f,t) \mid }{M_2(f,t) / \mid M_1(f,t)  \mid}\\
\end{cases}
\end{equation}
More precisely, the input of the network is $\mathbf{x} = [ILD,$ $\operatorname{Re}(IPD)], \operatorname{Im}(IPD)]$, where $\operatorname{Re}$ and $\operatorname{Im}$ denote real and imaginary part operators, respectively. Note that for the IPD, the frequency $f=0$ is discarded because it is constant for every observation. In general, the mapping between $V$ and the proposed feature is not unique. In particular, this happen when $\tau_2^1 = \tau_1^1$. In order to avoid this, we preventively pruned all the entries with $| \tau_2^1 - \tau_1^1 | < 10^{-6}$ from the dataset.

% The performances of cross-correlation methods (e.g. GCC-PHAT) depends on the extraction on local maxima. However in an reverberant scenario many local extrema at periodic intervals can be observed due to the reflections. So which peaks corresponds to the desired TDOA, iTDOA and TDOE? To avid this ambiguity advance methods can actually retrieve TDOAs and iTDOAs however it is challenging for them to estimate TDOE, which is an essential variable for defining an MIRAGE setup~\ref{fig:correlation}.
%The network is a simple multi-layers neural network (DNN). It has $D$ inputs corresponding to the dimension of the input feature vector $\mathbf{x}$ and $L = 3$ output nodes corresponding to the three target variables of $V$.
We use a simple fully-connected DNN architecture consisting of a $D$-dimensional input layer, a $3$-dimensional output layer, and 3 fully connected hidden layers with respective input sizes $500$, $300$ and $50$. Rectified linear unit (ReLU) activation functions are used except at the output layer, and each hidden layer has a dropout probability $p_\text{do} = 0.3$. We use the mean square error loss function for training and the Adam optimizer \cite{kingma2014adam}. The normalized root mean square error (nRMSE) is taken as validation metric\footnote{The nRMSE takes values between $0$ (perfect fit) and $\infty$ (bad fit). If it is equal to $1$, then the prediction is no better than a constant.}. The network is manually tuned on a validation set to find the best combination of number of hidden layers, their sizes and $p_\text{do}$. 
Once time delay estimates $\hat{V}$ are returned by the DNN, they are converted to synthetic local angular spectra and passed to $\Psi_\text{SRP}$ (See Sec. \ref{subsec:2D-SSL}) together with the relative positions of true and image microphones which are assumed known. We call this algorithm MIRAGE. The synthetic local angular spectra consist of Gaussians centered at $\hat{V}$ and with variances equal to the prediction errors made by the DNN on the validation set.

\section{Implementation and Results}\label{sec:exp}
To the best of the authors' knowledge, no reference implementation of algorithms for 2D-SSL using only 2 microphones is available to date. To check the validity of TDOA estimation, it is compared to GCC-PHAT using the true microphones (see Sec. \ref{subsec:1D-SSL}). For training and validation of the DNN we generate many random shoe-box room configurations using the software presented in \cite{Schimmel2009}. This software implements both the image-method for simulating reflections and a ray-tracing algorithm for diffusion. Room widths are uniformly drawn at random in $[3, 9]$ m, heights in $[2, 4]$ m. Random source/microphones positions and absorption coefficients for the 6 surfaces are used, respecting the close-surface scenario. In particular, the microphones are at most $30$ cm from the close-surface, placed $10$ cm from each other, the absorption coefficients of the other walls are uniformly sampled in $(0.5, 1)$ and the one of the close-surface is in $(0, 0.5)$. The same realistic diffusion profile \cite{gaultier2017vast} is used for all surfaces. Around $90,000$ audio scenes are generated this way, yielding reverberation times ($RT_{60})$ between $20$ ms and $250$ ms. 
%\textbf{TODO: compute DRR}
For training and validation, the RIRs are convolved with 1 sec of white-noise (wn) with no additional noise.
All signals and RIRs are sampled at $16$ kHz. The STFT is performed on $1024$ point with $50\%$ overlap. Finally the features are computed as in~\eqref{eq:features} yielding a vector of size $D = 1534$ for each observation.
While we validate the DNN on a portion of the dataset in a \textit{holdout} fashion, the test is conducted on 200 new RIRs convolved with both wn and speech (sp) utterances. This set is generated similarly to the training and validation sets. Moreover the recordings are perturbed by external white noise at 10 dB SNR (wn+n, sp+n). The speech signals are normalized speech utterances of various lengths (from $1$ s to $6$ s), randomly selected from the TIMIT corpus.
A free and open-source Matlab implementation of SRP-PHAT\footnote{\url{http://bass-db.gforge.inria.fr/bss_locate/}} is used to aggregate local angular spectra obtained from the DNN's output.
% The same toolbox is used for the implementation of SPR-PHAT with GCC-PHAT. For the latter method only real pairs are used.
A sphere sampling with $\ang{0.5}$ resolution and coordinates $\theta \in [-179, 180]$ and $\phi \in [0, 90]$ is used for the DOA search.

\setlength{\tabcolsep}{4.25pt}
\begin{table}[ht!]
\centering
\footnotesize
%\scriptsize
\begin{tabular}{cl|ccc|cc}
\toprule
           &                  &                  & nRMSE         &                   &\multicolumn{2}{c}{ACCURACY}  \\
		   & Input             &    \scriptsize{TDOA}  		 &   \scriptsize{iTDOA} 		 &     \scriptsize{TDOE} 		 & $\theta<\ang{10}$ &  $\theta<\ang{20}$ \\
\midrule
MIRAGE &   wn            &    0.18 & 0.28 & 0.25 	 & 4.10 (77)		   &  5.97 (97) \\
MIRAGE &   wn+n     &    0.68 & 0.69          &    0.89 			 & 5.00 (26)		   &  9.89 (54) \\
MIRAGE &  sp         &    0.31 & 0.34 &    0.56  & 4.83 (63)		   &  7.26 (82) \\
MIRAGE &  sp+n  &    0.99 & 0.98   	     &    1.48 			 & 4.60 (16)		   &  9.88 (35) \\
 GCC-PHAT   &   wn      &    0.21 		 &     -  		 & -				 & 4.22 (81) &   6.19 (97) \\
GCC-PHAT   &   wn+n            &    0.68 &     -  		 &       -			 & 4.03 (65) &   5.34 (83) \\
 GCC-PHAT   &  sp 		  &    0.32 		 &     -  		 &       -			 & 4.08 (82) &   5.34 (97) \\
 GCC-PHAT   &  sp+n   &    1.38 		 &     -  		 &       -			 & 4.70 (19) &   8.38 (32) \\
\bottomrule
\end{tabular}
\caption{Normalize root mean squared error for TDOA estimation and mean angular error in ${}^\circ$ (with accuracies ($\%$)) for AOA estimation with $\ang{10}$ and $\ang{20}$ angular tolerance.\vspace{-3mm}}
\label{tab:tdoas-aoa}
\end{table}

\setlength{\tabcolsep}{6pt}
\begin{table}[ht]
\footnotesize
\centering
\begin{tabular}{cl|cc|cc}
\toprule
 \textbf{DoA}      &               &  \multicolumn{2}{c|}{ACCURACY} &   \multicolumn{2}{c}{ACCURACY} \\
                   &               &  \multicolumn{2}{c|}{$<\ang{10}$} &   \multicolumn{2}{c}{$<\ang{20}$} \\
                   &    Input    &  $\theta$ &  $\phi$ &  $\theta$ &  $\phi$ \\
\midrule
 MIRAGE &  wn           &   4.5 (59) &  3.9 (71) &   6.8 (79) &   5.9 (88) \\
MIRAGE &  wn+n     &   4.4 (18) &  5.5 (26) &   9.4 (35) &  11.1 (66) \\
MIRAGE &  sp       &   4.6 (45) &  4.8 (59) &   8.1 (71) &   7.2 (83) \\
MIRAGE &  sp+n &   5.2 (17) &  5.9 (12) &  10.7 (38) &  12.3 (43) \\
\bottomrule
\end{tabular}
\caption{Mean angular error in ${}^\circ$ (with accuracies ($\%$)) for 2D SSL (azimuth and elevation) with $\ang{10}$ and $\ang{20}$ tolerance.\vspace{-5mm}}
\label{tab:doa}
\end{table}

%\section{Results and Discussion}
% In order to evaluate the performance three different metrics are used: first we compare TDOA in term of nRMSE for both $GCC-PHAT$ and $DNN$; second, we compare these two approaches for AOA estimation, that is the azimuth in the plane of the 2 real microphones, in term of accuracy, namely the percentage of angles correctly estimated above a certain threshold ($\ang{10}, \ang{20}$). Finally we present the fully 2D DoA estimation for both azimuth and elevation with the same metrics.

TDOA estimation errors using the proposed approach and GCC-PHAT are presented in Table~\ref{tab:tdoas-aoa}. Training a DNN to estimate TDOAs brings similar performances as GCC-PHAT in terms of nRMSE. Estimation of iTDOA and TDOE seems to be a harder task for the simple DNN we used. Nevertheless, our results confirm the possibility of retrieving early echoes from only two-microphone recordings. When some external noise is added, performance of both methods severely degrades. This is a well-know and expected behaviour for GCC-PHAT. It suggests that noise should be considered in the training phase of MIRAGE. When we compare the performance in terms of AOA, the two methods yield the same accuracy within a $\ang{20}$ threshold, as can be see in Table~\ref{tab:tdoas-aoa}. When a smaller tolerance is considered, GCC-PHAT outdoes the proposed approach in accuracy, with comparable errors.
%This behaviour is due to two aspects: first, the synthetic angular spectrum is a too simple model; second, since nRMSE was chosen as validation metrics, accuracy is not directly optimized. 
Again, when adding noise, performance decreases. In Table~\ref{tab:doa} the performance of the full 2D-SSL pipeline is showed. Within a tolerance of $\ang{20}$, the MIRAGE model allows estimation of both azimuth and elevation of the target source. However since in our data the 2 microphones were free to move, the inclinations of the true and image pairs are rarely flat. While this helps elevation estimation, it reduces the accuracy of predicting the right azimuth. While external noise is again decreasing the accuracy dramatically, it is interesting to notice that our DNN model trained and validated with white noise sources somewhat generalizes to speech sources.

\vspace{-2mm}
\section{Conclusion}
\vspace{-2mm}
In this paper we demonstrated how a simple echo model could allow 2D SSL with only two microphones, using simulated data. Future research will focus on extending this proof-of-concept to real data. The problem of echo-delay estimation proved to be very challenging, and extensions of the proposed learning scheme will be developed to obtain more reliable estimations of angular spectra. Extensions of the method to better handle various types of noise and emitted signals will also be sought. Finally, applications of the MIRAGE framework to larger microphone arrays, higher order echoes and a variety of tasks beyond SSL will be explored.

%\vfill
%\pagebreak

\bibliographystyle{IEEEbib}
\bibliography{refs}

\begin{thebibliography}{10}

\bibitem{rascon2017localization}
Caleb Rascon and Ivan Meza,
\newblock ``Localization of sound sources in robotics: A review,''
\newblock {\em Robotics and Autonomous Systems}, vol. 96, pp. 184--210, 2017.

\bibitem{Argentieri2015}
S.~Argentieri, P.~Dan{\`{e}}s, and P.~Sou{\`{e}}res,
\newblock ``{A survey on sound source localization in robotics: From binaural
  to array processing methods},''
\newblock {\em Computer Speech {\&} Language}, vol. 34, no. 1, pp. 87--112, nov
  2015.

\bibitem{Knapp1976}
C.~Knapp and G.~Carter,
\newblock ``{The generalized correlation method for estimation of time
  delay},''
\newblock {\em IEEE Transactions on Acoustics, Speech, and Signal Processing},
  vol. 24, no. 4, pp. 320--327, aug 1976.

\bibitem{DiBiase2001}
Joseph~H. DiBiase, Harvey~F. Silverman, and Michael~S. Brandstein,
\newblock ``{Robust Localization in Reverberant Rooms},''
\newblock in {\em Microphone Arrays: Signal Processing Techniques and
  Applications}, pp. 157--180. Springer, Berlin, Heidelberg, 2001.

\bibitem{Lebarbenchon2018}
Romain Lebarbenchon, Ewen Camberlein, Diego Carlo, Antoine Deleforge, and Nancy
  Bertin,
\newblock ``{Evaluation of an open-source implementation of the {SRP-PHAT}
  algorithm within the 2018 {LOCATA} challenge},''
\newblock in {\em 2018 IEEE-AASP Challenge on Acoustic Source Localization and
  Tracking (LOCATA), International Workshop on Acoustic Signal Enhancement},
  2018, pp. 2--3.

\bibitem{Scheuing2006}
Jan Scheuing and Bin Yang,
\newblock ``Disambiguation of tdoa estimates in multi-path multi-source
  environments (datemm).,''
\newblock in {\em ICASSP (4)}, 2006, pp. 837--840.

\bibitem{deleforge2015acoustic}
Antoine Deleforge, Florence Forbes, and Radu Horaud,
\newblock ``Acoustic space learning for sound-source separation and
  localization on binaural manifolds,''
\newblock {\em International Journal of Neural Systems}, vol. 25, no. 01, pp.
  1440003, 2015.

\bibitem{Vesperini2016}
Fabio Vesperini, Paolo Vecchiotti, Emanuele Principi, Stefano Squartini, and
  Francesco Piazza,
\newblock ``{A neural network based algorithm for speaker localization in a
  multi-room environment},''
\newblock in {\em 2016 IEEE 26th International Workshop on Machine Learning for
  Signal Processing (MLSP)}. Sep. 2016, pp. 1--6, IEEE.

\bibitem{Adavanne2017}
Sharath Adavanne, Archontis Politis, and Tuomas Virtanen,
\newblock ``Direction of arrival estimation for multiple sound sources using
  convolutional recurrent neural network,''
\newblock {\em CoRR}, vol. abs/1710.10059, 2017.

\bibitem{Perotin2018}
Laur{\'{e}}line Perotin, Romain Serizel, Emmanuel Vincent, and Alexandre
  Gu{\'{e}}rin,
\newblock ``{CRNN-based multiple DoA estimation using Ambisonics acoustic
  intensity features},''
\newblock {\em IEEE Journal of Selected Topics in Signal Processing
  \textit{(submitted)}}, 2018.

\bibitem{gaultier2017vast}
Cl{\'e}ment Gaultier, Saurabh Kataria, and Antoine Deleforge,
\newblock ``{VAST : The Virtual Acoustic Space Traveler Dataset},''
\newblock in {\em {International Conference on Latent Variable Analysis and
  Signal Separation (LVA/ICA)}}, Grenoble, France, Feb. 2017.

\bibitem{Nakashima2010}
Hiromichi Nakashima, Mitsuru Kawamoto, and Toshiharu Mukai,
\newblock ``{A localization method for multiple sound sources by using
  coherence function},''
\newblock {\em European Signal Processing Conference}, vol. 1, no. 3, pp.
  130--134, 2010.

\bibitem{dokmanic2013acoustic}
Ivan Dokmani{\'c}, Reza Parhizkar, Andreas Walther, Yue~M Lu, and Martin
  Vetterli,
\newblock ``Acoustic echoes reveal room shape,''
\newblock {\em Proceedings of the National Academy of Sciences}, vol. 110, no.
  30, pp. 12186--12191, 2013.

\bibitem{An2018}
Inkyu An, Myungbae Son, Dinesh Manocha, and Sung-Eui Yoon,
\newblock ``{Reflection-Aware Sound Source Localization},''
\newblock in {\em 2018 IEEE International Conference on Robotics and Automation
  (ICRA)}. may 2018, pp. 66--73, IEEE.

\bibitem{flanagan1993spatially}
James~L Flanagan, Arun~C Surendran, and Ea-Ee Jan,
\newblock ``Spatially selective sound capture for speech and audio
  processing,''
\newblock {\em Speech Communication}, vol. 13, no. 1-2, pp. 207--222, 1993.

\bibitem{dokmanic2015raking}
Ivan Dokmani{\'c}, Robin Scheibler, and Martin Vetterli,
\newblock ``Raking the cocktail party,''
\newblock {\em IEEE journal of selected topics in signal processing}, vol. 9,
  no. 5, pp. 825--836, 2015.

\bibitem{Scheibler2017}
Robin Scheibler, Diego~Di Carlo, Antoine Deleforge, and Ivan Dokmanic,
\newblock ``Separake: Source separation with a little help from echoes,''
\newblock in {\em 2018 {IEEE} International Conference on Acoustics, Speech and
  Signal Processing, {ICASSP} 2018, Calgary, Canada, Apr. 15-20}, 2018, pp.
  6897--6901.

\bibitem{Blandin2012}
Charles Blandin, Alexey Ozerov, and Emmanuel Vincent,
\newblock ``{Multi-source TDOA estimation in reverberant audio using angular
  spectra and clustering},''
\newblock {\em Signal Processing}, vol. 92, no. 8, pp. 1950--1960, 2012.

\bibitem{kingma2014adam}
Diederik~P. Kingma and Jimmy Ba,
\newblock ``Adam: {A} method for stochastic optimization,''
\newblock {\em CoRR}, vol. abs/1412.6980, 2014.

\bibitem{Schimmel2009}
Steven~M Schimmel, Martin~F Muller, and Norbert Dillier,
\newblock ``A fast and accurate “shoebox” room acoustics simulator,''
\newblock in {\em IEEE International Conference on Acoustics, Speech and Signal
  Processing, ICASSP 2009}, 2009, pp. 241--244.

\end{thebibliography}

\end{document}